\documentclass[twocolumn,aps,prl,superscriptaddress,showpacs,amsmath,amssymb,floats]{revtex4}

\usepackage{amsmath}
\usepackage{amssymb}
\usepackage{graphicx}
\usepackage{keyval}
\usepackage{dcolumn}
\usepackage{bm}
\usepackage{color}
\usepackage{txfonts}
\usepackage[]{sidecap}

\begin{document}

\title{Electronic structure of the BaTi$_2$As$_2$O parent compound of the titanium based oxypnictide superconductor}

\author{H. C. Xu}

\author{M. Xu}

\author{R. Peng}

\author{Y. Zhang}

\author{Q. Q. Ge}

\author{F. Qin}

\author{M. Xia}
\affiliation{State Key Laboratory of Surface Physics, Department of Physics, and
Advanced Materials Laboratory, Fudan University, Shanghai 200433,
People's Republic of China}

\author{J. J. Ying}

\author{X. H. Chen}
\affiliation{Hefei National Laboratory for Physical Science at Microscale and Department of Physics, University of Science
and Technology of China, Hefei, Anhui 230026, People¡¯s Republic of China}

\author{M. Arita}

\author{K. Shimada}

\author{M. Taniguchi}

\affiliation{Hiroshima Synchrotron Radiation Center and Graduate School of Science, Hiroshima University, Hiroshima 739-8526, Japan}

\author{D. H. Lu}

\affiliation{Department of Applied Physics and Stanford Synchrotron Radiation Laboratory, Stanford University, Stanford, California 94305, USA}

\author{B. P. Xie}
\email{bpxie@fudan.edu.cn}

\author{D. L. Feng}
\email{dlfeng@fudan.edu.cn}

\affiliation{State Key Laboratory of Surface Physics, Department of Physics, and
Advanced Materials Laboratory, Fudan University, Shanghai 200433,
People's Republic of China}

\date{\today}
\begin{abstract}

The electronic structure of BaTi$_2$As$_2$O, a parent compound of the newly discovered titanium-based oxypnictide superconductors, is studied by angle-resolved photoemission spectroscopy. The electronic structure shows multi-orbital nature and possible three-dimensional character. An anomalous temperature-dependent spectral weight redistribution and broad lineshape indicate the incoherent nature of the spectral function. At the density-wave-like transition temperature around 200~K, a partial gap opens at the Fermi patches. These findings suggest that BaTi$_2$As$_2$O is likely a charge density wave material in the strong interaction regime.

\end{abstract}


\maketitle

\section{Introduction}

Planar square lattice made of $3d$ transition elements and anions has been proven to be a rich playground for high temperature superconductivity, as exemplified in the CuO$_2$ plane of the cuprate high temperature superconductors, and the FeAs or FeSe plane of the iron-based superconductors. Recently, superconductivity has been discovered in Ba$_{1-x}$Na$_x$Ti$_2$Sb$_2$O ($0.0{\leq}x{\leq}0.33$) \cite{Yajima2012,Doan2012}, a  titanium-based oxypnictide with an anti-CuO$_2$ type Ti$_2$O plane and Sb above and below the center of each Ti$_2$O square. From the structural point of view, such a lattice is an intriguing combination of both CuO$_2$ and FeAs type of lattices, thus it may be another family of unconventional superconductors based on titanium, although the maximal $T_c$ is currently only 5.5~K upon Na doping \cite{Doan2012}.

The signature of possible charge density wave or spin density wave has been observed in  Ba$_{1-x}$Na$_x$Ti$_2$Sb$_2$O and other compounds with (Ti$_2$Pn$_2$O)$^{2-}$ (Pn=As, Sb) layers \cite{Axtell1997,Ozawa2004,Liu2009,Liu2010,Wang2010}, resembling that in the iron-based high temperature superconductors. The instability of the ordered state, such as spin fluctuation, often plays an important role in the unconventional superconductivity \cite{Singh2012}. Therefore,  it is critical to reveal the nature of the density wave transition. Theoretical calculations suggest that the nested Fermi surface sections drive the density wave instability \cite{Biani1998,Pickett1998,Yan2013}. However, the experimental electronic structure of such materials has not been reported. Moreover, density wave order has not been identified in powder neutron diffraction experiments on Na$_2$Ti$_2$Pn$_2$O (Pn=As, Sb) \cite{Ozawa2000,Ozawa2001}. Only recently have the muon spin rotation and nuclear magnetic resonance studies excluded magnetic order in Ba$_{1-x}$Na$_x$Ti$_2$Sb$_2$O and favor a charge density wave picture \cite{Kitagawa2013,Rohr2013}.

\begin{figure}[bt]
\includegraphics[width=8.7cm]{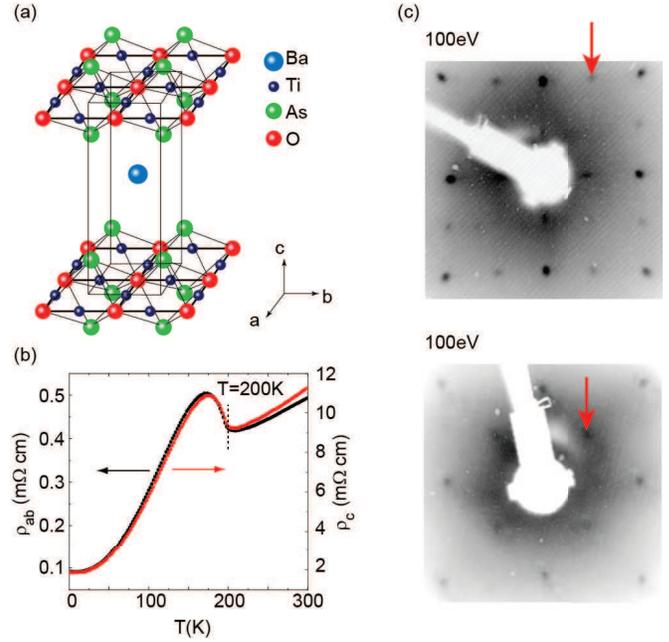}
\caption{(color online). (a) The crystal structure of BaTi$_2$As$_2$O. (b) In-plane and out-of-plane resistivity as a function of temperature for BaTi$_2$As$_2$O. (c) The low-energy electron diffraction patterns of two BaTi$_2$As$_2$O samples. The arrows mark the reconstruction spots.}
\label{crystal}
\end{figure}

\begin{figure}[bt]
\includegraphics[width=8.7cm]{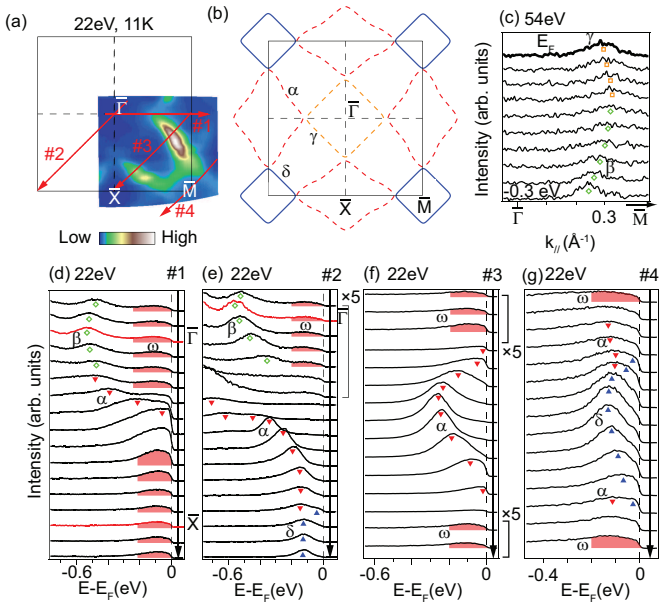}
\caption{(color online). Electronic structure of BaTi$_2$As$_2$O measured at 11~K. (a) Photoemission intensity map at the Fermi energy ($E_F$) integrated over an energy window of $\pm$10~meV. $\bar{\Gamma}$, \={M}, and  \={X} stand for high symmetric points of the projected two-dimensional Brillioun zone (BZ). (b) The Fermi surface sheets obtained by tracking the Fermi crossings in panel (a). The dashed lines indicate the hole pockets, and the solid blue lines indicate the electron pockets.  (c) Momentum distribution curves (MDC's) taken with $p$ polarized light along  $\bar{\Gamma}$-\={M} near zone center. (d)-(g) Energy distribution curves (EDC's) taken with circular polarized light at 11~K along cuts \#1, \#2, \#3, and \#4 in panel (a), respectively. Some parts of the spectra are magnified for a better view.}
\label{Bands}
\end{figure}

BaTi$_2$As$_2$O is an isostructural compound of the superconducting BaTi$_2$Sb$_2$O, with isovalent substitution of Sb by As [Fig.~\ref{crystal}(a)] \cite{Wang2010}. Thus it could be viewed as a parent compound of the titanium-based superconductors. A density wave transition takes place around 200~K, as shown by the resistivity data [Fig.~\ref{crystal}(b)]. In this Article, we investigate the electronic structure of BaTi$_2$As$_2$O with angle resolved photoemission spectroscopy (ARPES). The photoemission spectra show broad lineshape without sharp quasipartical peak near the Fermi energy ($E_F$). The spectral weight at E$_F$ forms Fermi pockets with parallel sections and Fermi patches (finite spectral weight away from Fermi surfaces). Our polarization and photon energy dependent studies reveal the multi-orbital and three-dimensional nature of this material. Intriguingly, the spectral weight redistributes over a large energy scale as a function of temperature, indicating the incoherent nature of the spectral function. Furthermore, a partial gap opens at the Fermi patches around the density wave transition temperature, which is similar to charge density wave materials in the strong interaction regime \cite{Shen2007,Shen2008}.

\section{Experimental}

Single crystals of BaTi$_2$As$_2$O were grown by flux method \cite{Wang2010,Wang2009}. The ARPES data were taken at the Beamline~1 and Beamline~9 of Hiroshima synchrotron radiation center (HiSOR). Temperature dependent study was conducted at Beamline~5-4 of Stanford Synchrotron Radiation Lightsource (SSRL). The 7~eV laser data were taken with an in-house setup at Fudan University. All data were collected with Scienta R4000 electron analyzers. The energy resolution is 20~meV at HiSOR Beamline~1, 10~meV at HiSOR Beamline~9, 5~meV at SSRL, and 4~meV for the laser setup. The samples were cleaved \emph{in-situ}, and measured under ultrahigh vacuum better than $3\times10^{-11}$~mbar. The low energy electron diffraction (LEED) patterns in Fig.~\ref{crystal}(c) indicate the square lattice of cleaved surface [the $ab$ plane in Fig.~\ref{crystal}(a)]. Additional surface reconstruction spots, as indicated by the arrows, appear depending on cleavages and probably come from the reconstruction of the half-layer Ba atoms at the surface similar to those in BaFe$_{2-x}$Co$_x$As$_2$ \cite{Massee2009}. Aging effects were strictly monitored during the experiments.

\section{Band structure}

Figure~\ref{Bands}(a) shows the photoemission intensity map of BaTi$_2$As$_2$O at $E_F$. As sketched in Fig.~\ref{Bands}(b), the Fermi surfaces in the projected two-dimensional Brillioun zone (BZ) consists of one square shaped hole pocket ($\gamma$) around the $\bar{\Gamma}$ point, four diamond shaped hole pockets ($\alpha$) centered at four \={X} points, and four square shaped electron pockets ($\delta$) centered at four \={M} points. The square shaped pockets at $\bar{\Gamma}$ and \={M} show multiple parallel sections, providing possible Fermi surface nesting condition for density wave instabilities, as suggested in previous theoretical calculations \cite{Biani1998,Pickett1998,Singh2012,Yan2013}. At the first glance, the Fermi surfaces may come from the warping and hybridization of two sets of quasi-1D sections along $(1,1)$ and $(-1,1)$ directions \cite{Biani1998}. In this case, the Fermi pocket at the zone center should be electron-like; however in contrast to this, the square shaped pocket formed by $\gamma$ is a hole pocket. As shown by the momentum distribution curves in Fig.~\ref{Bands}(c), an electron-like band $\beta$ fades out below $E_F$, while a hole-like band $\gamma$ appears and crosses $E_F$ near the zone center.

Figure~\ref{Bands}(d) shows the energy distribution curves (EDC's) along $\bar{\Gamma}$-\={X}, crossing the hole pockets centered at $\bar{\Gamma}$ and \={X} [cut \#1 in Fig.~\ref{Bands}(a)]. Band $\alpha$ contributes prominent spectral weight and forms the hole pocket around \={X}. Along cut \#2, band $\alpha$ disperses towards $E_F$ and coincides with the bottom of band $\delta$, which forms the electron pocket at zone corner [Fig.~\ref{Bands}(e)]. Bands $\alpha$ and $\delta$ keep broad lineshape while approaching $E_F$, and cross $E_F$ without a sharp quasi-particle peak. Along both cut \#1 and \#2, band $\beta$ is observed below $E_F$ near zone center [Figs.~\ref{Bands}(d)-\ref{Bands}(e)], while the band $\gamma$, which forms the hole pocket at $\bar{\Gamma}$, could hardly be tracked in EDC's [Figs.~\ref{Bands}(d)-\ref{Bands}(e)] due to its weak intensity and nearly vertical dispersion [Fig.~\ref{Bands}(c)]. Figure~\ref{Bands}(f) shows the dispersion of band $\alpha$ along cut \#3 between two neighboring hole pockets. In Fig.~\ref{Bands}(g), bands $\alpha$ and $\delta$ along cut \#4 show similar dispersive structures with those near zone corner along cut \#2.
Moreover, in Figs.~\ref{Bands}(d)-\ref{Bands}(g), there is spectral weight with broad lineshape (shaded area) between -0.2~eV and $E_F$, labeled as $\omega$, contributing finite spectral weight over almost the entire BZ, which resembles the Fermi patches observed in 2H-Na$_x$TaS$_2$ before \cite{Shen2007}, which is a charge density wave material with strong electron-phonon interactions.

\begin{figure}[bt]
\includegraphics[width=8.7cm]{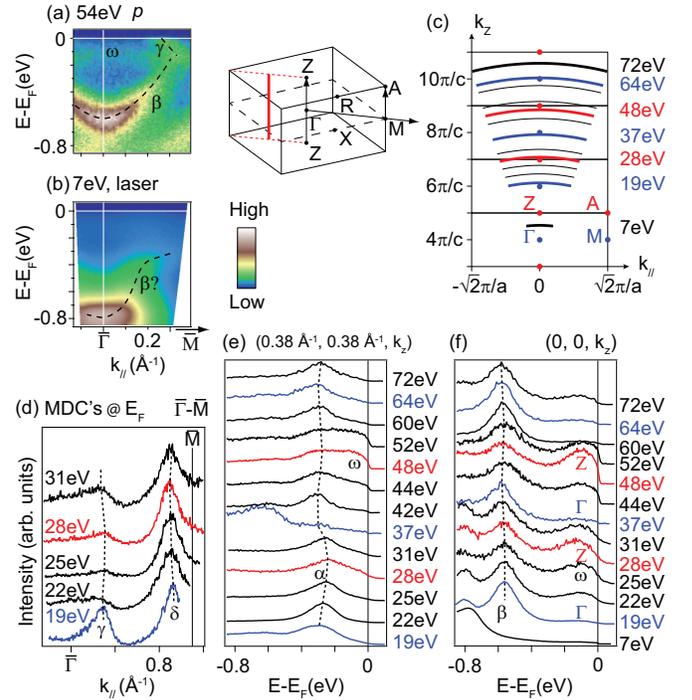}
\caption{(color online). (a) Spectral intensity along  $\bar{\Gamma}$-\={M} near zone center. (b) is the same as panel (a), but taken with 7~eV laser. (c) Sketch of BZ's in $\Gamma$-M-A-Z plane. The corresponding position for different photon energies are indicated, with the inner potential assumed to be 12~eV. The inset illustrates the three-dimensional BZ of BaTi$_2$As$_2$O. (d) MDC's at $E_F$ along $\bar{\Gamma}$-\={M} as a function of photon energy. (e) EDC's taken along (0.38$\AA^{-1}$, 0.38$\AA^{-1}$, $k_z$) (the red solid line in the three-dimensional BZ) with various photon energies as shown in panel (c). (f) is the same as panel (e), but along (0, 0, $k_z$).}
\label{PhDep}
\end{figure}

The band dispersion along $k_z$ is investigated by photon energy dependent study. In Figs.~\ref{PhDep}(a)-\ref{PhDep}(b), the photoemission spectra taken with 7~eV laser and 54~eV photon energy show distinct band structures. This implies strong dispersion along $k_z$, consistent with the previous calculations on its sibling materials \cite{Pickett1998,Singh2012,Yan2013}. We have performed a detailed study with photon energies ranging from 19 to 72~eV, covering more than two BZ's along $k_z$ [Fig.~\ref{PhDep}(c)]. The dispersions of bands $\gamma$, $\delta$ [Fig.~\ref{PhDep}(d)], and $\alpha$ [Fig.~\ref{PhDep}(e)] match the period of BZ along $k_z$ if assuming the inner potential to be 12~eV. Moreover, the broad feature $\omega$ shows a variation of intensity that matches the period of BZ [Figs.~\ref{PhDep}(e)-\ref{PhDep}(f)]. On the other hand, as shown in Figs.~\ref{PhDep}(d)-\ref{PhDep}(f), the dispersion is rather weak above 19~eV. Such a weak dispersion along $k_z$ contradicts the distinct spectra taken with 7~eV laser, suggesting a loss of dispersive information along $k_z$ probably due to the poor $k_z$ resolution for photon energy range [19~eV, 72~eV]. In the universal curve of escaping depth \textit{vs.} kinetic energy, the mean escaping depth $\xi$ of the photoelectron excited by photons at this energy range is around 7~\AA \cite{Hufner}, yielding a ${\Delta}k_z\approx2\pi/\xi$ as large as the BZ size. This would largely smear out the dispersive information along $k_z$ and result in a weak dispersion and/or Fermi patches with periodic variation of intensity, as indeed observed in our data. The escaping depth increases rapidly with photon energies from 19~eV to 7~eV by more than three times \cite{Hufner}, providing a much better $k_z$ resolution, which could explain the distinct spectra taken with 7~eV laser.
Intriguingly, the spectra taken with 7~eV photon energy still show intrinsic broad features.

\begin{figure}[bt]
\includegraphics[width=8.7cm]{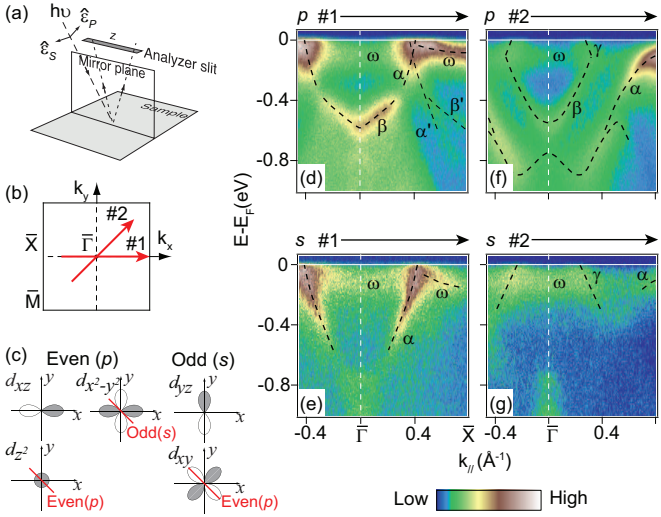}
\caption{(color online). (a) Experimental setup for polarization-dependent ARPES. (b) The sketch of BZ and cuts. (c) The $3d$ orbitals are classified into even and odd symmetries with respect to the $\Gamma$-X-R-Z plane. The spacial symmetries with respect to $\Gamma$-M-A-Z plane are illustrated in red color. The geometries indicated in the parentheses provide finite photoemission intensity. (d)-(e) Photoemission intensity along cut \#1 taken with 48~eV photon energy in $s$-polarized geometry, and in $p$-polarized geometry, respectively. (f)-(g) are the same as panels (d)-(e), but taken along cut \#2.} \label{Orbital}
\end{figure}

\section{Multi-orbital nature of bands near E$_F$}

For compounds with (Ti$_2$Pn$_2$O)$^{2-}$ (Pn=As, Sb) layers, the density of states near $E_F$ is dominated by Ti $3d$ electrons according to theoretical calculations \cite{Pickett1998,Singh2012,Yan2013}, and there is only one $3d$ electron for each Ti$^{3+}$ ion. However, our data show multiple bands crossing $E_F$ in BaTi$_2$As$_2$O, suggesting multi-orbital composition of these bands. To reveal the orbital composition near $E_F$, polarization-dependent ARPES study was conducted in both $p$ and $s$ geometries. Figure~\ref{Orbital}(a) shows the geometries of the photoemission experiment setup, where the path of the incident photon beam and the sample surface normal define the mirror plane. The matrix element of photoemission could be described by
\[|M^{\textbf{k}}_{f,i}|^2=|\langle\phi^{\textbf{k}}_{f}|\varepsilon \cdot \textbf{r}|\phi^{\textbf{k}}_{i}\rangle|^2\]
where the $\phi^{\textbf{k}}_{f}$ and $\phi^{\textbf{k}}_{i}$ are final- and initial-states, respectively \cite{Damascelli2003}. Given that the final state is a plane wave and its symmetry is always even against the mirror plane, the photoemission intensity would be suppressed if the initial state and the $(\hat{\varepsilon} \cdot \textbf{r})$ have opposite symmetries. With the k$_x$ and k$_y$ directions in Fig.~\ref{Orbital}(b) defined as the Ti-As-Ti direction, the five $d$ orbitals are classified into even and odd symmetries in Fig.~\ref{Orbital}(c) with respect to the $\Gamma$-X-R-Z plane. The symmetries with respect to the $\Gamma$-M-A-Z plane are indicated in red [Fig.~\ref{Orbital}(c)]. In the current experimental setup, the even (odd)  orbitals should be visible in $p$ ($s$) geometry [Fig.~\ref{Orbital}(c)]. In addition, for the out-of-plane component of the $p$ polarized light, the signal of orbitals extending out-of-plane would be enhanced in this geometry.

As illustrated in Fig.~\ref{Orbital}(b), polarization-dependent ARPES study around $\bar{\Gamma}$ was conducted along both $\bar{\Gamma}$-\={X} (\#1) and $\bar{\Gamma}$-\={M} (\#2) with 48~eV photons. In Fig.~\ref{Orbital}(d), some features folded from band $\alpha$ and $\beta$ show up around \={X} (noted as $\alpha'$ and $\beta'$), consistent with the surface reconstruction in the upper panel of Fig.~\ref{crystal}(c). In Figs.~\ref{Orbital}(d)-\ref{Orbital}(e), the band $\alpha$ is visible in both geometries along $\bar{\Gamma}$-\={X}, indicating its mixed orbital composition with both odd and even symmetries. The spectral weight of feature $\omega$ is enhanced near \={X} in $p$ geometry, indicating orbital with even symmetry or out-of-plane component. At other momenta, $\omega$ is visible in both geometries, suggesting mixed orbital composition.

Along $\bar{\Gamma}$-\={M}, the band $\alpha$ is visible in $p$ geometry [Fig.~\ref{Orbital}(f)] and suppressed in $s$ geometry [Fig.~\ref{Orbital}(g)], suggesting its even composition along this direction. In Figs.~\ref{Orbital}(f)-\ref{Orbital}(g), band $\gamma$ and $\omega$ are visible in both geometries, indicating their mixed orbital composition. Band $\beta$ and other features at higher binding energies show even symmetry in both $\bar{\Gamma}$-\={X} and $\bar{\Gamma}$-\={M} direction [Figs.~\ref{Orbital}(d)-\ref{Orbital}(g)].

\begin{figure}[bt]
\includegraphics[width=8.7cm]{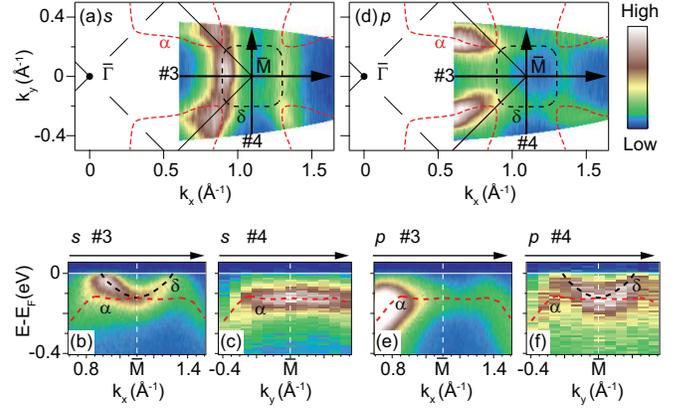}
\caption{(color online). (a) The photoemission intensity map around \={M} taken with 22~eV photon energy in $s$-polarized geometry. (b) and (c) show the photoemission intensity along cut \#3 and \#4 in panel (a), respectively. (d)-(f) are the same as panels (a)-(c), but taken with $p$-polarized geometry. The spectra along cut \#4 are generated by combining individual EDC's at corresponding momenta.} \label{OrbitalsM}
\end{figure}

As shown by the Fermi surface mapping in $s$ geometry [Fig.~\ref{OrbitalsM}(a)], the top and bottom edges of $\delta$ pocket is suppressed , while its left and right edges are fairly intensive. Below $E_F$ in $s$ geometry, the band $\delta$ is visible along cut \#3 but not visible along cut \#4 [Figs.~\ref{OrbitalsM}(b)-\ref{OrbitalsM}(c)]. In $p$ geometry, the band $\delta$ is not visible along cut \#3 but visible along cut \#4 [Figs.~\ref{OrbitalsM}(e)-\ref{OrbitalsM}(f)]. This suggests that the orbital composition of band $\delta$ is odd along cut \#3 and even along cut \#4, \textit{i.e.}, a switch of orbital symmetries under four fold rotation about \={M}.

Around \={M}, band $\alpha$ shows flat dispersion and is visible in both geometries and along both cuts \#3 and \#4 [Figs.~\ref{OrbitalsM}(b)-\ref{OrbitalsM}(c) and \ref{OrbitalsM}(e)-\ref{OrbitalsM}(f)], indicating mixed orbital composition. A change of orbital composition between the dispersive part and the flat part could be observed from the intensity change of band $\alpha$ in Figs.~\ref{OrbitalsM}(c) and \ref{OrbitalsM}(e).

To determine the orbital composition, orbital resolved band structure calculation is required. However, there is no band structure calculation of BaTi$_2$As$_2$O yet. Nevertheless, our results show that the features near $E_F$ come from the hybridization of multiple Ti 3$d$ orbitals. The calculation on BaTi$_2$Sb$_2$O \cite{Singh2012} show muti-orbital compostion of the bands near $E_F$, which supports our findings. The multi-orbital character in BaTi$_2$As$_2$O resembles that in the iron-based superconductors \cite{Singh2008,Kuroki2008,Graser2009}.

\begin{figure}[bt]
\includegraphics[width=8.6cm]{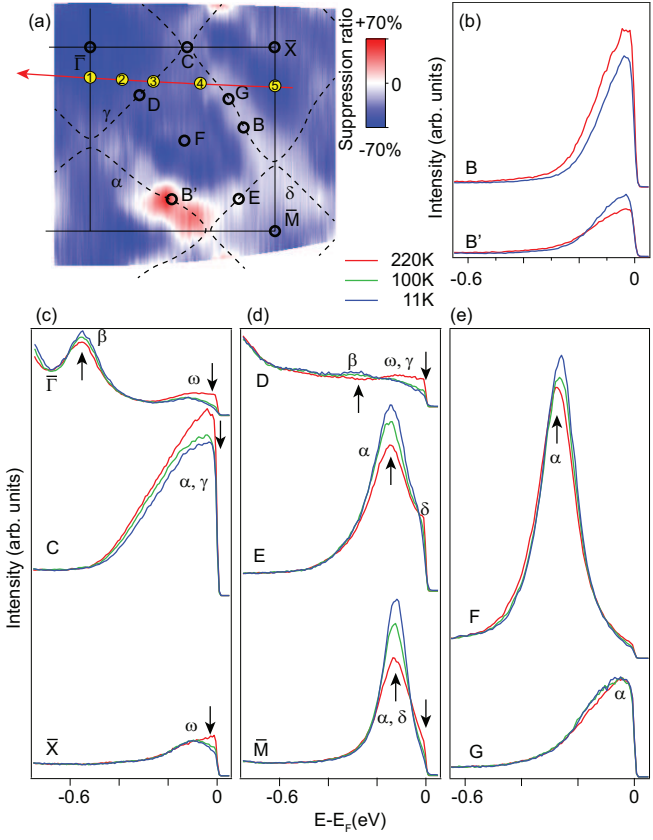}
\caption{(color online). (a) The false color plot of the spectral weight suppression ratio from 220~K to 11~K, \textit{i.e.},  $[I_{E_F}(11K)-I_{E_F}(220K)])/I_{E_F}(220K)$. The spectra are integrated over [$E_F$-25~meV, $E_F$+5~meV] after dividing the resolution convolved Fermi-Dirac function. (b)-(e) Temperature dependence of EDC's at various momenta as marked in panel (a). The thermal broadening near $E_F$ are removed by dividing the spectra with resolution convolved Fermi-Dirac function at corresponding temperatures and then multiplied by that of 11~K, \textit{i.e.}, $I(E,k,T)/\tilde{f}_{FD}(E,T){\times}\tilde{f}_{FD}(E,11K)$, while $\tilde{f}_{FD}(E,T)$ stands for the resolution convolved Fermi-Dirac function. The panels (a) and (b) are from the same data set, while the panels (c)-(e) are from individual temperature-dependent studies along fixed momentum cuts. Photoemission data were taken with 22~eV photons.} \label{TDep}
\end{figure}

\section{Temperature dependent study}

Now we turn to the temperature-dependent study on the density wave transition in BaTi$_2$As$_2$O. In the conventional picture of density wave transition, a gap opens along parallel Fermi surface sheets when the system enters the density wave state \cite{Pickett1998,Singh2012,Yan2013}. Considering the Fermi patches spreading over the BZ and the lack of sharp Fermi crossings, we investigated the temperature dependence over a quarter of the BZ.
Figure~\ref{TDep}(a) shows the false color map of the spectral weight suppression ratio at $E_F$ from 220~K to 11~K, \textit{i.e.},  $(I_{E_F}(11K)-I_{E_F}(220K))/I_{E_F}(220K)$. The spectral weight is suppressed at the Fermi patches away from Fermi crossings, where the broad feature $\omega$ dominates. The parallel Fermi surface sheets formed by bands $\gamma$ and $\delta$, where density wave instability was suggested, do not show prominent suppression.
Most part of the $\alpha$ pocket shows weak suppression, while one branch at the lower side of BZ shows great enhancement, as exemplified by comparing the temperature dependence of EDC's at momenta B and B' [Fig.~\ref{TDep}(b)]. The partial suppression of spectral weight at $E_F$ is consistent with the metallic behavior below transition [Fig.~\ref{crystal}(b)], and agrees with previous optical study on its sibling material Na$_2$Ti$_2$As$_2$O \cite{Huang2013}. However, such an asymmetric suppression/enhancement for the $\alpha$ pocket cannot be explained by matrix element effect, possibly due to other momentum dependent effects.

Figures~\ref{TDep}(c)-\ref{TDep}(e) show the EDC's at various momenta as marked in Fig.~\ref{TDep}(a). The EDC's are divide by the resolution convolved Fermi-Dirac function at corresponding temperature and multiplied by that of 11~K to remove thermal broadening effect near $E_F$. The suppression of spectral weight takes place near $E_F$, as indicated by the down arrows on the EDC's for momenta $\bar{\Gamma}$, C, \={X} [Fig.~\ref{TDep}(c)], D, and \={M} [Fig.~\ref{TDep}(d)].
On the other hand, the spectral weight of bands $\alpha$, $\beta$, and $\delta$ below $E_F$ enhances with decreasing temperature [up arrows in Figs.~\ref{TDep}(c)-\ref{TDep}(e)].
Thus the evolution of spectral weight distribution with temperature could be viewed as a spectral weight redistribution from Fermi patches at $E_F$ to the dispersive bands at higher binding energies, which saves energy. The energy scale of the spectral weight redistribution ranges from 0.1~eV for bands $\alpha/\delta$ at zone corner, to 0.6~eV for band $\beta$ at zone center, which is far beyond the energy range of thermal effect and is not expected in a non-interacting electron system. The spectral weight enhancement for bands $\alpha$, $\beta$, and $\delta$ do not slow down below 100~K [up arrows in Figs.~\ref{TDep}(c)-\ref{TDep}(e)], indicating that it is not relevant to the density wave transition around 200~K. Similar large scale spectral weight redistribution was observed in Sr$_2$CuO$_2$Cl$_2$, which is explained by multiple initial-/final-states induced by strong coupling between electrons and bosons \cite{Kim2002}. The large scale spectral weight redistribution in BaTi$_2$As$_2$O could originate from similar strong coupling effect, which is probably momentum dependent and induces an asymmetric suppression/enhancement as observed in Fig.~\ref{TDep}(a). On the other hand, the spectral weight suppression at $E_F$ mostly takes place between 220~K and 100~K [down arrows in Figs.~\ref{TDep}(c)-\ref{TDep}(e)], suggesting density wave transition related component.

\begin{figure}[bt]
\includegraphics[width=8.6cm]{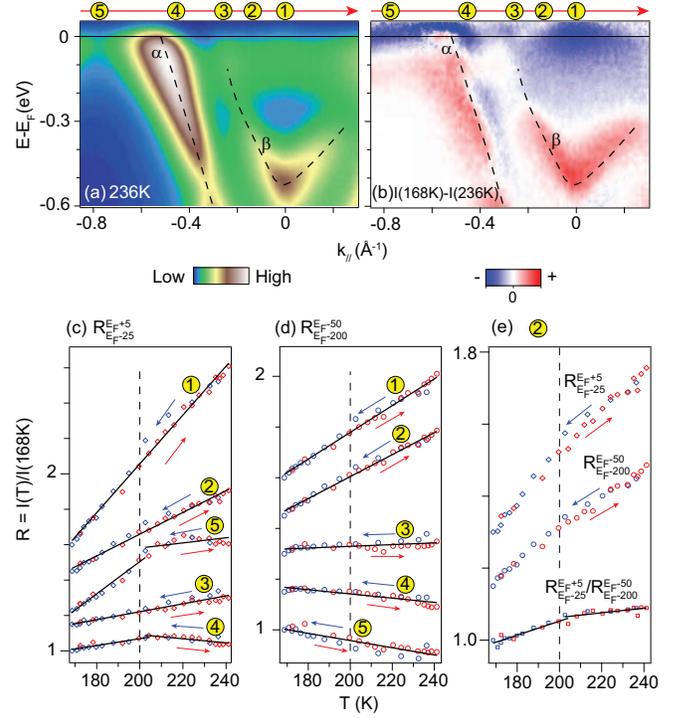}
\caption{(color online). (a) Photoemission spectra taken at 236~K with 22~eV photons along cut \textcircled{\footnotesize{1}}-\textcircled{\footnotesize{5}}, as marked in Fig.~\ref{TDep}(a). (b) the difference between the spectra at 236~K and 168~K, which are divided by the resolution convoluted Fermi-Dirac distributions at the corresponding temperatures before subtraction. (c) Density of states as a function of temperature normalized by that at 168~K, integrated over [$E_F$-25~meV, $E_F$+5~meV] ($R_{E_F-25}^{E_F+5}$) at corresponding momenta \textcircled{\footnotesize{1}}-\textcircled{\footnotesize{5}}. The blue data points and arrows indicate process of decreasing temperature, while the red ones stand for increasing temperature. (d) is the same as panel (c), but integrated over [$E_F$-200~meV, $E_F$-50~meV] ($R_{E_F-200}^{E_F-50}$). (f) $R_{E_F-25}^{E_F+5}$, $R_{E_F-200}^{E_F-50}$ at momentum \textcircled{\footnotesize{2}}, and their ratio ($R_{E_F-25}^{E_F+5}$/$R_{E_F-200}^{E_F-50}$) as a function of temperature. The data in (c)-(e) are vertically offset for a better view.} \label{TDep2}
\end{figure}

In order to investigate the density wave transition related component, we conduct detailed temperature-dependent study around 200~K along the cut through momenta \textcircled{\footnotesize{1}}$\sim$\textcircled{\footnotesize{5}} as marked in Fig.~\ref{TDep}(a). This cut is selected to cover momenta \textcircled{\footnotesize{1}} and \textcircled{\footnotesize{2}} (the Fermi patch inside the Fermi surface $\gamma$), momentum \textcircled{\footnotesize{5}} (the Fermi patch inside the Fermi surface $\alpha$ ), and momentum \textcircled{\footnotesize{4}} (the Fermi patch between the Fermi surfaces), while the bands $\alpha$ and $\beta$ are resolved [Fig.~\ref{TDep2}(a)]. Figure~\ref{TDep2}(b) shows the spectral weight difference between 236~K and 168~K. The major suppression of spectral weight takes place around $E_F$, while the enhancement is mainly at higher binding energies on bands $\alpha$ and $\beta$,
demonstrating the spectral weight redistribution. Detailed temperature evolution of relative spectral weight at momenta \textcircled{\footnotesize{1}}$\sim$\textcircled{\footnotesize{5}} are plotted in Figs.~\ref{TDep2}(c)-\ref{TDep2}(d), for energy window [$E_F$-25~meV, $E_F$+5~meV] ($R_{E_F-25}^{E_F+5}$), and [$E_F$-200~meV, $E_F$-50~meV] ($R_{E_F-200}^{E_F-50}$), respectively. At most momenta and binding energies, the spectral weight show smooth suppression/enhancement though 200~K, demonstrating the spectral weight redistribution independent of the density wave transition. On the other hand, slope changes are observed around 200~K for momentum \textcircled{\footnotesize{4}} and \textcircled{\footnotesize{5}} [Fig.~\ref{TDep2}(c)], indicating gap opening of density wave transition at these momentum.
At momentum \textcircled{\footnotesize{2}}, after dividing the temperature-dependent curve of $R_{E_F-25}^{E_F+5}$ by that of $R_{E_F-200}^{E_F-50}$, both of which show smooth evolution, a slope change around 200~K is readily observable [$R_{E_F-25}^{E_F+5}$/$R_{E_F-200}^{E_F-50}$ in Fig.~\ref{TDep2}(e)], demonstrating a partial gap opening buried under the large spectral weight redistribution. Note that the density wave transition related slope changes take place at Fermi patches (e.g., momenta \textcircled{\footnotesize{2}}, \textcircled{\footnotesize{4}}, and \textcircled{\footnotesize{5}}) without well-defined Fermi surfaces, suggesting that the partial gap opening takes place at Fermi patches, rather than at the nested Fermi surface sections based on the conventional density wave picture \cite{Pickett1998,Singh2012,Yan2013}.

\section{Discussions and conclusion}

The dispersive features in photoemission spectra of BaTi$_2$As$_2$O show broad lineshape and cross $E_F$ without sharp quasiparticle peaks. On the other hand, the overall bandwidth show little renormalization compared with the calculated band structure of sibling materials \cite{Singh2012,Yan2013}. These characters resemble those in polaronic systems like La$_{1.2}$Sr$_{1.8}$Mn$_{2}$O$_{7}$, where incoherent broad features dominate the photoemission spectra with vanishingly small quasiparticle weight, while following the bare band dispersion \cite{Mannella2005,Dessau1998}. The polaronic picture is further supported by the large scale temperature-dependent spectral weight redistribution, which resembles that in Sr$_2$CuO$_2$Cl$_2$ induced by strong electron correlations with phonons or magnetic excitations \cite{Kim2002}.

Due to the interference of large scale temperature-dependent spectral weight redistribution, it is hard to extract the exact momentum distribution of density wave gap opening. However, the signature of partial gap opening on Fermi patches rather than Fermi surfaces provides some clue for further investigation. Since the Fermi patches partly come from the convolving of highly $k_z$-dispersive bands, an out-of-plane nesting vector could be expected to form the density wave instability, like that in VSe$_2$ \cite{Strocov2012}. On the other hand, charge density wave instability formed by Fermi patches or barely occupied states were reported in 2H-Na$_x$TaS$_2$ and NbSe$_2$ \cite{Shen2007,Shen2008}, which show polaronic signatures. Therefore, BaTi$_2$As$_2$O is likely a similar charge density wave material in the strong interaction regime. Moreover, considering the conventional s-wave superconductivity in 2H-Na$_x$TaS$_2$ and NbSe$_2$ and recent reports \cite{Kitagawa2013,Rohr2013,Subedi2013}, the superconductivity in titanium oxypnictides could also be conventional.

In summary, we have carried out a systematic ARPES study on BaTi$_2$As$_2$O and revealed its band structure with multi-orbital nature and possible three-dimensional character. The spectral weight at $E_F$ forms both Fermi surfaces and Fermi patches. Broad lineshape and anomalous temperature dependence of spectral weight redistribution suggest the incoherent nature of the spectral functions due to strong interactions. Partial gap opening at the transition temperature are observed on Fermi patches rather than the parallel sections of Fermi surfaces, which resembles charge density wave materials in the strong interaction regime.

\section{Acknowledgements}

We gratefully acknowledge the helpful discussions with Prof. J. P. Hu. This work is supported in part by the National Science Foundation of China, and the National Basic Research Program of China (973 Program) under the grant Nos.~2012CB921400, 2011CB921802, 2011CBA00112, 2011CB309703, 91026016. SSRL is operated by the US DOE, BES, Divisions of Chemical Sciences and Material Sciences.

\end{document}